\documentclass[a4paper,11pt]{article}
\usepackage{jcappub} 

\usepackage{derivative}
\usepackage{mathrsfs}
\usepackage{physics}
\usepackage{bm} 
\usepackage{color}
\usepackage{xcolor}
\usepackage[normalem]{ulem}

\arxivnumber{1234.56789} 
\title{\boldmath Primordial Black Hole Formation from the Upward Step Model: Avoiding Overproduction}

\author[a]{Xiaoding Wang,} 

\author[b,c,d,e]{Xiao-Han Ma,} 

\author[b,c,d]{Yi-Fu Cai} 

\affiliation[a]{The Key Laboratory of Cosmic Rays (Tibet University), Ministry of Education, Lhasa 850000, Tibet, China}

\affiliation[b]{Department of Astronomy, School of Physical Sciences, University of Science and Technology of China, Hefei, Anhui 230026, China}

\affiliation[c]{CAS Key Laboratory for Research in Galaxies and Cosmology, School of Astronomy and Space Science, University of Science and Technology of China, Hefei, Anhui 230026, China}

\affiliation[d]{Deep Space Exploration Laboratory, Hefei 230088, China}

\affiliation[e]{Kavli Institute for the Physics and Mathematics of the Universe (WPI), UTIAS The University of Tokyo, Kashiwa, Chiba 277-8583, Japan}

\emailAdd{mxh171554@mail.ustc.edu.cn}

\abstract{
We investigate the formation of primordial black holes (PBHs) in an upward step inflationary model, where nonlinearities between curvature perturbations and field fluctuations introduce a cutoff, deviating from the Gaussian case. This necessitates a reevaluation of PBH formation, as $\mathcal{R}$ is not the optimal variable for estimating abundance. Using the extended Press-Schechter formalism, we show that non-Gaussianity modifies both the curvature perturbation profile $\mathcal{R}(r)$ and the integration path in probability space, significantly impacting PBH abundance. 
Our results reveal that the abundance initially increases with the parameter $h$, which characterizes the relaxation stage after the step. However, beyond a critical value ($h \simeq 5.9$), it sharply declines before rising again. Furthermore, we demonstrate that non-Gaussianity introduces uncertainties in indirect PBH observations via gravitational waves. Notably, we present an example where a positive $f_{\rm NL}$ does not necessarily enhance PBH production, contrary to conventional expectations. 
Finally, by accounting for non-perturbative effects, we resolve the overproduction of PBHs suggested by pulsar timing array (PTA) data, underscoring the critical importance of incorporating non-Gaussianity in future studies.

}

\begin{document}

\maketitle
\flushbottom

\section{Introduction}\label{sec:intro}
PBHs have become one of the most compelling topics, captivating the attention of cosmologists and astrophysicists for over a decade, as their existence offers a natural explanation for numerous mysteries in modern cosmology. 
Unlike black holes formed from stellar collapse, PBHs originate in the primordial universe and result from the collapse of disturbed primordial radiation fields \cite{Hawking:1971ei,Carr:1974nx,Carr:1975qj}. They decoupled from baryons in the very early universe and thus can be considered candidates for dark matter \cite{Khlopov:2008qy,Belotsky:2014kca,Carr:2020xqk}. Furthermore, PBHs of various masses can seed the formation of supermassive black holes (SMBHs) observed by the James Webb Space Telescope (JWST) in the early universe \cite{Dolgov:2023ijt}, directly or indirectly source the gravitational waves detected by LIGO/VIRGO \cite{LIGOScientific:2016aoc,LIGOScientific:2016sjg,LIGOScientific:2017bnn,LIGOScientific:2017ycc,LIGOScientific:2017vox}, and PTAs collaborations (NANOGrav \cite{NANOGrav:2023gor,NANOGrav:2023hde}, EPTA \cite{EPTA:2023fyk,EPTA:2023sfo,EPTA:2023xxk}, PPTA \cite{Reardon:2023gzh,Zic:2023gta,Reardon:2023zen} and CPTA \cite{Xu:2023wog}) and even heat our universe through Hawking radiation before big bang nucleosynthesis (BBN) \cite{He:2022wwy}. This capability is due to the fact that the mass of PBHs depends both on the size of the Hubble horizon at the time of their formation and on primordial fluctuations. Consequently, PBHs are able to surpass the mass limits typically associated with astrophysically formed black holes.

Since PBHs form from the collapse of the overdense regions in the very early universe, the statistics of primordial fluctuations and the criteria for collapse determine both their initial abundance and initial mass function. These are crucial properties that allow us to search for PBHs. On the other hand, detecting PBHs also serves as a smoking gun, providing insights into the physics behind primordial fluctuations and revealing the unknown history of the early universe.

From the latest cosmic microwave background (CMB) observations, we know that the comoving curvature perturbvation $\mathcal{R}$ is nearly scale invariant, quasi-adabatic and Gaussian on large scales \cite{Planck:2018vyg}.
Specifically,  $\mathcal{P}_{\mathcal{R}}(k) = A_s(k/k_{\text{pivot}})^{n_s - 1} $  with $A_s \simeq 2 \times 10^{-9}$,  $n_s \simeq 0.96$ and $k_{\text{pivot}} = 0.05~\text{Mpc}^{-1}$. If perturbations at smaller scales exhibit the same characteristics, the chances of PBH formation in our (observable) universe are minimal. However, various indications suggest that the power spectrum of curvature perturbations was enhanced at smaller scales, leading to a rich set of phenomenological effects, including the formation of a detectable fraction of PBHs and the generation of scalar-induced gravitational waves (SIGWs).

Although the observation of a stochastic gravitational wave background in the $\text{nHz}$ band by PTAs in 2023 can be interpreted as gravitational waves from a population of supermassive black hole binaries \cite{Ellis:2023dgf,Sato-Polito:2023gym,Xiao:2024nmi,Chen:2024hqc,Toubiana:2024bil}, the possibility that this signal originates from the primordial epoch and has a cosmological source remains an open question. An enhanced primordial power spectrum with $\mathcal{P}_{\mathcal{R}} \sim 10^{-1}$ has also been proposed as a potential explanation \cite{NANOGrav:2023hvm,Franciolini:2023pbf,Ellis:2023oxs}.
Subsequently, it was realized that if all gravitational wave signals were induced by enhanced scalar perturbations, PBHs produced simultaneously should also have been observed, yet they remain undetected in current direct observations. This results in the PBH overproduction problem, which may suggest the non-Gaussianity of primordial perturbations \cite{Franciolini:2023pbf,Ellis:2023oxs}.

Instead of using the curvature perturbation on comoving slices $\mathcal{R}$ or the density contrast $\delta\rho/\rho$, the compaction function $\mathcal{C}(r,t,\vb{x}) \equiv 2\delta M(r,t,\vb{x})/L(r,t)$, where $L(r,t)$ is the areal radius, is now commonly employed as the appropriate variable for calculating the abundance of PBHs \cite{Harada:2015yda,Musco:2018rwt,Young:2022phe,Harada:2023ffo,Franciolini:2023pbf,Ferrante:2022mui,Gow:2022jfb}.
Following the same philosophy as the Press-Schechter formalism used to calculate the halo mass function, the extended Press-Schechter formalism calculates the abundance of PBHs by integrating the probability density function (PDF) $\mathbb{P}(\mathcal{C})$ from the threshold. In the Gaussian scenario, all necessary information for the PDF is encapsulated within the power spectrum $\mathcal{P}_{\mathcal{R}}$. This dependency is critical for constraining PBHs through indirect observations, including SIGWs.

Since there are infinitely many ways to deviate from Gaussian statistics, accurately characterizing such deviations becomes crucial when considering non-Gaussianity. 
Researchers use non-Gaussianity parameters to discuss local non-Gaussianity, such as $f_{\rm{NL}}$ and $g_{\rm{NL}}$, based on a perturbative series of the random variable:
\begin{equation}\label{eq:difinition of local NG}
\begin{aligned}
\mathcal{R}(\vb*{x}) = \mathcal{R}_{\rm G}(\vb*{x}) + \frac{3}{5}f_{\rm{NL}}\left[ \mathcal{R}^2_{\rm G}(\vb*{x}) - \langle \mathcal{R}^2_{\rm G}(\vb*{x}) \rangle \right] + g_{\rm{NL}}\mathcal{R}^3_{\rm G}(\vb*{x}) + \cdots,
\end{aligned}
\end{equation}
where $\mathcal{R}_{\rm G}(\vb*{x})$ denotes the Gaussian random field. 
This expression provides a model-independent description of local non-Gaussianity and is widely adopted in the literature. Using this perturbative but model-independent approach, several studies have revealed the non-Gaussian effects on SIGWs and PBHs \cite{Cai:2018dig,Ferrante:2023bgz,Franciolini:2023pbf,Liu:2023ymk,Choudhury:2023fwk,Iovino:2024tyg,Choudhury:2024kjj}.
However, this perturbative approach may not be effective when there is a significant non-Gaussian tail. Specifically, when calculating the abundance of PBHs, we are analyzing statistics related to rare events described by the tail of the PDF. Recently, several works have investigated the formation of PBHs in the presence of a non-Gaussian tail \cite{Franciolini:2018vbk,Panagopoulos:2019ail,Figueroa:2020jkf,Achucarro:2021pdh,Cai:2022erk,Chen:2022dqr,Pi:2022ysn}. 
Analysis of the latest 15-year dataset released by PTAs in 2023 indicated that non-Gaussian primordial perturbations with a positive $f_{\rm NL}$ tend to exacerbate the PBH overproduction problem \cite{Nakama:2015nea,Dandoy:2023jot,Franciolini:2023pbf,Inomata:2023zup,Cai:2023dls,Wang:2023ost,Liu:2023ymk,Lewicki:2024ghw}, leading to the disfavoring of several models, including those with an upward step feature in the inflaton potential \cite{Firouzjahi:2023xke}.

However, a positive $f_{\rm NL}$ does not always enhance the abundance of PBHs when non-perturbative features play a role in the tail of the PDF. In models with an upward step in the inflaton potential, there exists a cutoff in the PDF of curvature perturbations \cite{Cai:2022erk}, which can significantly boost or suppress the PBH abundance. In this work, we demonstrate that upward step models can still potentially produce the stochastic gravitational wave background (SGWB)  observed by PTAs without the overproduction of PBHs, even when $f_{\rm NL}$ is positive, by employing the extended Press-Schechter formalism. This simple example shows how the abundance of PBHs can deviate greatly from calculations based on the primordial scalar power spectrum with corrections from $f_{\rm NL}$.

The paper is organized as follows: In Section \ref{sec:revisiting_up_step}, we revisit the single-field inflation model with an upward step in the inflaton potential and analyze the non-Gaussian tail of the curvature perturbation using the $\delta N$ formalism. In Section \ref{sec:PBH formation in upward step model}, we set up the extended Press-Schechter formalism and estimate PBH formation in the slow-roll-step-slow-roll model. In Section \ref{sec_PTA_PBHs_overp}, we discuss the SIGWs accompanying PBH formation and reconsider the issue of overproduction by accounting for non-perturbative non-Gaussianity. Section \ref{sec:conclusion} is devoted to the conclusion.

\section{Revisiting upward step model}\label{sec:revisiting_up_step}
In a standard single-field inflation scenario, the slow-roll (SR) phase is necessary to produce a nearly scale-invariant power spectrum $\mathcal{P}_{\mathcal{R}_{\rm G}}$ at CMB scales. The slow-roll conditions are:
\begin{align}\label{eq: slow-roll condition}
    \varepsilon_{ H}\equiv-\frac{\dot{H}}{H^2} \ll 1~, \qquad \left|\eta_{ H}\right|\equiv \left|\frac{\dot{\varepsilon_{ H}}}{H \varepsilon_{ H}}\right| \ll 1 ~,
\end{align}
which ensure that inflationary expansion lasts long enough. In models with a canonical kinetic term, $K \propto \partial_{\mu}\phi\partial^{\mu}\phi$, the slow-roll condition is typically interpreted as the requirement for a sufficiently flat potential $V(\phi)$.  We use ` $\dot{}$ ' to denote the derivative with respect to comoving cosmic time $t$.
However, PBH formation often demands non-trivial features in $V(\phi)$, such as an inflection point or step-like features 
\cite{Covi:2006ci,Hamann:2007pa,Mortonson:2009qv,Adshead:2011jq,Miranda:2013wxa,Miranda:2015cea,Mishra:2019pzq,Kefala:2020xsx,Inomata:2021uqj,Dalianis:2021iig,Wolfson:2021zsw,Inomata:2021tpx,Cai:2021zsp,Cai:2022erk,Animali:2022otk,Pi:2022ysn,Kawaguchi:2023mgk,Wang:2024wxq}. These non-trivial features cause the inflaton to deviate from the slow-roll phase, thereby enhancing the scalar power spectrum and generating non-Gaussianity on specific scales.
To discuss the non-perturbative non-Gaussianity (i.e., the non-Gaussian tail) in the upward step model and its application to PBH formation in light of 15-year PTAs observations, we revisit the upward step model here. We consider a piecewise potential joined by an upward step, with a height $\Delta V$, parameterized as follows,
\begin{equation}\label{eq:simplest upward potential}
	\begin{aligned}
		V(\phi) = &V_0\left[1 + \sqrt{2\epsilon_I}\left(\phi - \phi_{\star}\right)\right]\Theta\left(\phi - \phi_{\star}\right)\\ 
		&\qquad + \left(V_0 + \Delta V\right)\left[1 + \sqrt{2\epsilon_{II}}\left(\phi - \phi_\star\right) \right ]\Theta\left(\phi_\star - \phi\right)~,
	\end{aligned}
\end{equation}
where $\Theta(x)$ is the Heaviside step function\footnote{
In a realistic model, we reformulate the second line of \eqref{eq:simplest upward potential} as $+ V_0 + \left[\Delta V + (V_0+\Delta V) \sqrt{2\epsilon_{II}}(\phi - \phi_\star) \right] \Theta_{\rm S}(\phi_\star -\phi; \lambda)$, where $\Theta_{\rm S}(x = \phi_\star - \phi;\lambda)$ is a smoothed step function designed to ensure the continuity of $V(\phi)$ and its first derivative. The parameter $\lambda > 0$ controls the degree of smoothing, and in the limit $\lambda \to \infty$, $\Theta_{\rm S}(x;\lambda)$ reduces to the Heaviside step function $\Theta(x)$. It is defined as $\Theta_{\rm S}(x;\lambda) = \lambda x + \lambda x(\lambda x-1) \left[ 1-2\lambda x + (1-\lambda x)V_0\sqrt{2\epsilon_{I}}/(\lambda \Delta V ) \right]$ for $0 < \lambda x < 1$, with $\Theta_{\rm S}(x;\lambda) = 0$ for $\lambda x < 0$ and $\Theta_{\rm S}(x;\lambda) = 1$ for $\lambda x > 1$.
}

\begin{figure}[htbp]
    \centering
    \includegraphics[width=0.7\linewidth]{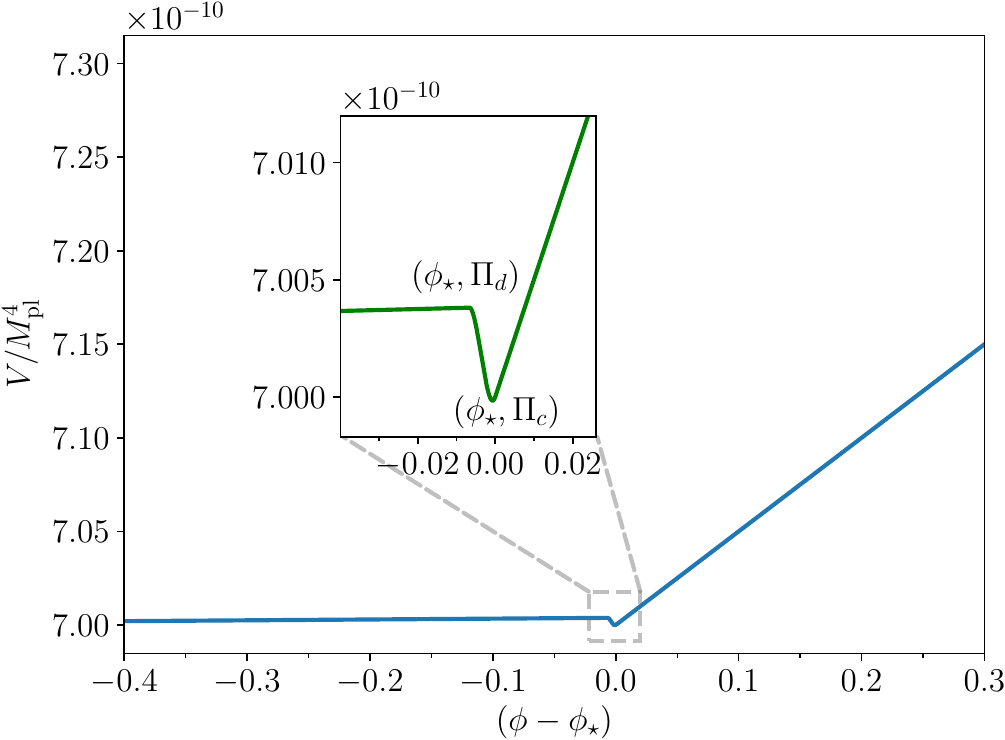}
    \caption{The inflationary potential $V(\phi)$ is given by \eqref{eq:simplest upward potential}. The parameters are chosen as  $V_0=7 \times 10^{-10} M^4_{\rm pl}$, $\Delta V = 3.85805 \times 10^{-13} M^4_{\rm pl}$, $\epsilon_{I} = 2.551 \times 10^{-3}$, $\epsilon_{II} = 4.077 \times 10^{-7}$, and $\lambda = 1.5 \times 10^2 $.}
    \label{fig:V_phi}
\end{figure}
The Friedmann equations and the equation of motion for the inflaton $\phi$ in the flat Friedmann-Lemaître-Robertson-Walker (FLRW) background are expressed as
\begin{align}
    &3M_{\mathrm{pl}}^2H^2=\frac{1}{2}\dot\phi^2+V(\phi),\\
    &M_{\mathrm{pl}}^2\dot H= -\frac{1}{2}\dot\phi^2,\\
    &\ddot\phi+3H\dot\phi+V_{,\phi}=0.
\end{align}
For simplicity, we use the convention that the reduced Planck mass $M_\mathrm{pl}\equiv \sqrt{1/8\pi G}=1$ to simplify the expressions. Here, $V_{,\phi}$ represents the derivative of $V(\phi)$ with respect to $\phi$. It is more convenient to rewrite the Klein-Gordon (K-G) equation using the e-folding number $\dd n =H\dd t$ as a time variable:
\begin{align}\label{eq: background EOM for phi}
       &\frac{\dd^2\phi}{\dd n^2}+(3-\varepsilon_H)\frac{\dd \phi}{\dd n}+\frac{V_{,\phi}}{H^2}=0.
\end{align}
In this model with a canonical kinetic term, $\Delta V/V$ is usually small to ensure a healthy end to inflation (whereas for non-canonical cases, the inflaton can overcome much steeper barriers \cite{Cai:2023uhc}). 
Thus, when examining the dynamics near the sharp upward step, $H$ can be considered constant, leading to a nonlinear relationship between the field velocity before and after the step.
\begin{align}\label{eq:pid}
    \Pi_{\rm d} \simeq \sqrt{\Pi^2_{\rm c} - 6\frac{\Delta V}{V}} ~,
\end{align}
where the inflaton's momentum is defined as 
\begin{align}\label{eq: Pi}
    \Pi \equiv - \dv{\phi}{n} .
\end{align}


Subsititute the potential \eqref{eq:simplest upward potential} into K-G equation and ignore the $\varepsilon_H$ term, then we have
\begin{align}\label{eq:Background equation in delta N}
    \frac{\dd^2\phi}{\dd n^2}+3  \frac{\dd\phi}{\dd n}+3\sqrt{2\epsilon_{X}}=0,
\end{align}
which gives the solution:
\begin{align}\label{eq:trajectory}
    \phi(n) = C_{1,X} \mathrm{e}^{-3(n-n_{\star})} -\sqrt{2\epsilon_{X}} (n - n_\star)+ C_{2,X} ~,
\end{align}
where the coefficients $C_{1,X}$ and $C_{2,X}$ are determined by matching condition at $\phi_{\star}$. This solution suggests that the phase $(\phi(n), \Pi(n))$ of the system can be expressed as:
\begin{align}
    3\left[\phi(n) - \phi_{\star}\right] - \left[\Pi(n)-\Pi_{\rm c}\right] &= -3\sqrt{2\epsilon_{I}}(n - n_\star) ~, \quad \phi > \phi_{\star} ~,\label{eq:trajectorie 2}\\
    3\left[\phi(n) - \phi_{\star}\right] - \left[\Pi(n)-\Pi_{\rm d}\right] &= -3\sqrt{2\epsilon_{II}}(n - n_\star) ~, \quad \phi < \phi_{\star} ~.\label{eq:trajectorie 2_2}
\end{align}
The parameter $\phi_\star$, which determines the location of the step, is set by the model. Thus, in the classical regime (ignoring the diffusion noise from sub-horizon fluctuations, i.e., quantum diffusion, see \cite{Pattison:2017mbe,Animali:2024jiz,Vennin:2024yzl}), the dynamics of the second stage are fully governed by the phase at the top of the step $(\phi_{\star}, \Pi_{\rm d})$.
The sudden change in field velocity, as given by \eqref{eq:pid}, can push the inflaton significantly away from the attractor solution, with the subsequent relaxation leading to several non-trivial phenomena. Thus, it is convenient to define two dimensionless parameters here:
\begin{align}\label{eq:g and h}
    g \equiv \frac{\Pi_{\rm d}}{\Pi_{\rm c}} ~ ,\qquad h \equiv \frac{6\sqrt{2\epsilon_{II}}}{\Pi_{\rm d}} ~.
\end{align}
Here, $0 < g < 1$ quantifies the velocity loss due to the upward step, and $h > 0$ measures the deviation of the initial velocity $\Pi_{\rm d}$ from the slow-roll attractor in the second stage\footnote{Notably, this definition of $h$ differs from \cite{Cai:2022erk} as we have absorbed the negative sign into the definition of $\Pi$.}. 

The enhancement of the power spectrum of curvature perturbations was initially observed and later applied to generate PBHs. As discussed in \cite{Cai:2022erk}, by solving the Sasaki-Mukhanov equation with appropriate matching conditions at the step, the power spectrum of curvature perturbations $\mathcal{R}_{\rm{G},k}$ exiting the horizon near $\phi_{\rm c}$ is enhanced by a factor of $g^{-2}$, allowing it to easily meet the condition for producing PBHs as dark matter, $\mathcal{P}_{\mathcal{R}_{\rm G}} \simeq 10^{-2}$.

In addition to the amplitude and location of the peak, the growth behavior before the peak, as well as the IR and UV behavior of $\mathcal{P}_{\mathcal{R}_{\rm G}}(k)$, are discussed in detail. For clarity in the subsequent discussion, we summarize the key features of the power spectrum for the SR-upward-step-SR model here:

\begin{equation}\label{eq:power spectrum}
    \mathcal{P}_{\mathcal{R}_{\rm G}}(k) \simeq \left\{
    \begin{aligned}
         & \frac{1}{2\epsilon_{I}} \left(\frac{H}{2\pi}\right)^2\,; && \text{IR part} ~, & \\
        & k^4\,; && k^4 \text{ growth before the peak}~, & \\
        &\frac{1}{2g^2\epsilon_{II}} \left(\frac{H}{2\pi}\right)^2\,; && \text{peak }~, &\\
        &\frac{1}{2\epsilon_{II}} \left(\frac{H}{2\pi}\right)^2\,; &&\text{UV part }~.&
    \end{aligned}
    \right.
\end{equation}
A non-Gaussian tail was identified in this model through an analysis using the $\delta N$ formalism \cite{Cai:2022erk}. The $\delta N$ formalism connects the curvature perturbation $\mathcal{R}$ at the end of inflation to the quantum fluctuations of the inflaton by evaluating the e-folding number backward during the superhorizon evolution of a Hubble patch from the end of inflation. 

The $\delta N$ calculation can be expressed as
\begin{align}
    \delta N = N_{\text{tot}}(\phi_{\rm i}+ \delta \phi,\Pi_{\rm i} + \delta \Pi;\phi_{\rm end},\Pi_{\rm end}) - N_{\text{tot}}(\phi_{\rm i},\Pi_{\rm i};\phi_{\rm end},\Pi_{\rm end}) ~,
\end{align}
where $N_{\text{tot}} \equiv  n_{\rm end} -  n_{\rm i} > 0$ is counted backward to the end of inflation.

In the upward step model, a highly non-trivial non-Gaussian tail arises in the perturbation modes exiting the horizon near the step (i.e., $\phi \simeq \phi_{\star}$) due to the non-linear dependence of $\Pi(n)$ in \eqref{eq:pid}. For modes that crossed the horizon much earlier, the attractor behavior of \eqref{eq:trajectorie 2} suppresses the initial deviation of $\Pi(n)$ from the attractor solution before the step. Thus, $\delta N$ is primarily contributed by the linear part of \eqref{eq:trajectory},
\begin{align}
    \mathcal{R} = \delta N \simeq \frac{\delta \phi}{\sqrt{2\epsilon_{I}}}~, \quad\text{long wave length modes.}
\end{align}
Based on the same argument, modes that exit the horizon long after the step, as well as long before the end of inflation, also follow a linear $\delta N$ result:
\begin{align}
    \mathcal{R} \simeq \frac{\delta \phi}{\sqrt{2\epsilon_{II}}}~, \quad\text{short wavelength modes.}
\end{align}

However, for modes exiting the horizon just before the step, where $n_{\rm c} - n_{\rm i} \ll 1$, \eqref{eq:trajectorie 2} implies
\begin{align}\label{eq:Pi_c approxi}
    \delta \Pi_{\rm c} \simeq \delta \Pi_{\rm i} - 3 \delta \phi ~,
\end{align}
meaning that the fluctuation of $\phi_{\rm i}$ affects the second slow-roll stage nonlinearly through $\delta \Pi_{\rm d}(\delta \phi)$. For these modes, the e-folding number, counted backward from the end of inflation, is given by:
\begin{align}\label{eq:e-folds Ntot}
    N_{\text{tot}} = N_{I} + N_{II} = (n_{\star} - n_{\text{i}}) + (n_{\text{end}} - n_{\star}) ~.
\end{align}
Since $N_I$ and $N_{II}$ are evaluated using different solutions of \eqref{eq:Background equation in delta N}, we cannot cancel $n_{\star}$ between the two stages. Consequently, both terms contribute to the final $\delta N$ in principle.

Equations \eqref{eq:trajectorie 2} and \eqref{eq:trajectorie 2_2} suggest that $N_I$ and $N_{II}$ in \eqref{eq:e-folds Ntot} can be expressed as
\begin{align}
    N_I &\equiv n_{\star} - n_{\rm i} = \frac{1}{3\sqrt{2\epsilon_I}}\left[ 3\left(\phi_{\rm i} - \phi_{\star}\right)  -\left( \Pi_{\rm i} - \Pi_{\rm c} \right) \right] ~,\label{eq:unperturbed N1}\\
    N_{II} &\equiv  n_{\rm end} -  n_{\star} = \frac{1}{3\sqrt{2\epsilon_{II}}}\left[ \left( \Pi_{\rm end} - \Pi_{\rm d} \right) - 3\left(\phi_{\rm end} - \phi_{\star}\right) \right] ~. \label{eq:unperturbed N2}
\end{align}
From the expressions of $N_I$ and $N_{II}$ above, the calculation of $\delta N$ follows directly by subtracting \eqref{eq:e-folds Ntot} from its perturbed counterpart, where $\phi_{\rm i} \to \phi_{\rm i} + \delta \phi$, $\Pi_{\rm i} \to \Pi_{\rm i} + \delta \Pi$,  $\Pi_{\rm c} \to \Pi_{\rm c} + \delta \Pi_{\rm c}$, and $\Pi_{\rm d} \to \Pi_{\rm d} + \delta \Pi_{\rm d}$.

Equation \eqref{eq:Pi_c approxi} ensures that the contribution from $\delta N_I$ approximately vanishes. Furthermore, setting $\delta \Pi = 0$ in the adiabatic limit (see the detailed discussion in \citep{Wang:2024wxq}) leads to the final expression for $\delta N$.\footnote{One can easily verify that this result agrees with that of \cite{Cai:2022erk} in the linear piecewise potential limit, where the second derivative of the potential can be neglected.}

\begin{equation}
\begin{aligned}
    \mathcal{R} \simeq \delta N_{II} \simeq  - \frac{\Pi_{\rm d}}{3\sqrt{2\epsilon_{II}}}\left[ \sqrt{1 - \frac{6}{g}\frac{\delta\phi}{\Pi_{\rm d}} + 9\left(\frac{\delta\phi}{\Pi_{\rm d}}\right)^2} - 1\right] ~.
\end{aligned}
\end{equation}
Neglecting the quadratic term of the perturbation inside the squareroot allows us to express $\mathcal{R}$ in a more compact form:
\begin{equation}\label{eq:non_linear R}
	\mathcal{R} \simeq -\frac{2}{h} \left[\sqrt{1 - h\mathcal{R}_\mathrm{G}} - 1\right], \quad \text{where} \quad \mathcal{R}_\mathrm{G} \equiv \frac{6}{gh} \left(\frac{\delta\phi}{\Pi_{\rm d}}\right) ~.
\end{equation}
Here, $\mathcal{R}_\mathrm{G}$ represents the linear part of $\mathcal{R}$ and remains Gaussian if $\delta \phi$ follows a Gaussian distribution.

This nonlinear $\mathcal{R}$ results in a highly non-Gaussian tail in the PDF $\mathbb{P}[\mathcal{R}]$, which is challenging to achieve through perturbative methods. For example, when $g^2 \ll 1$, using the definition of the local non-Gaussianity parameter \eqref{eq:difinition of local NG}, $f_{\rm NL}$ in our model is given by:
\begin{align}
    f_{\rm NL} \simeq \frac{5}{12}h > 0 ~.
\end{align}
This result is also justified by applying the generalized consistency relation \cite{Firouzjahi:2023xke}. Since $f_{\rm NL}$ is the coefficient of the quadratic term, a positive $f_{\rm NL}$ increases the probability of a positive $\mathcal{R}$. From this perspective, it becomes clear that models with a positive $f_{\rm NL}$ are more likely to produce PBHs \cite{Firouzjahi:2023xke}.

However, \eqref{eq:non_linear R} imposes a cutoff on $\mathcal{R}$, meaning that the probability of $\mathcal{R} > 2/h$ is zero, which deviates significantly from the Gaussian case. This cutoff suggests that PBHs formation must be considered more carefully when the non-Gaussian tail plays a significant role.

\section{PBH formation in upward step model}\label{sec:PBH formation in upward step model}

\subsection{Basic setup for extended Press-Shecheter formalism}\label{sec:Basic setup for extended Press-Shecheter formalism}
The formation of PBHs is thought to occur due to the collapse of super-horizon fluctuations. Only exceptionally large fluctuations, which are rare, can lead to PBH formation upon reentering the horizon after inflation. Since the collapse occurs in real space and is governed by the non-linearity of gravity, the full shape of the fluctuations must be known to determine whether a PBH will form. It has been shown that in the presence of non-Gaussianity, all orders of the correlators of curvature perturbation $\expval{\mathcal{R}\mathcal{R}\cdots \mathcal{R}}$ are needed to describe the full profile of fluctuations \cite{Germani:2019zez,Ferrante:2022mui}. 
Compared to the curvature perturbation $\mathcal{R}$, the compaction function $\mathcal{C}$ \cite{Shibata:1999zs}, which is defined as the mass excess relative to the background, serves as a more accurate estimator for the threshold of PBH formation \cite{Young:2014ana,Harada:2015yda,Musco:2018rwt,DeLuca:2022rfz,Harada:2023ffo,Raatikainen:2023bzk}.
On super-horizon scales, assuming spherical symmetry, the compaction function $\mathcal{C}$ during the radiation-dominated era can be expressed in terms of the curvature perturbation $\mathcal{R}$
 in the comoving gauge\footnote{In the literature, the curvature perturbation $\zeta$, defined in the uniform density gauge, is often used when discussing the compaction function. Since $\zeta$ and $\mathcal{R}$ are equivalent on super-horizon scales, for simplicity in notation, we use $\mathcal{R}$ throughout this paper.}:
\begin{equation}\label{eq:compaction function linear}
	\mathcal{C}(r) = \frac{2}{3} [1 - (1 + r\mathcal{R}^{\prime}(r))^2] = \mathcal{C}_\ell - \frac{3}{8} \mathcal{C}_\ell^2 ~, \quad \text{where}\quad \mathcal{C}_\ell \equiv - \frac{4}{3}r\mathcal{R}^{\prime} ~.
\end{equation}
$\mathcal{C}_\ell $ is referred to as the linear compaction function, and `` $^\prime$ '' denotes the derivative with respect to $r$.

Numerical results \cite{Escriva:2019phb} show that the averaged compaction function $\bar{\mathcal{C}}$ within a radius $r_{\rm m}$ is a reliable indicator of the collapse leading to PBH formation,
\begin{align}
    \mathcal{\bar{C}}\equiv\frac{3}{L^{3}(r_m)}\int_0^{L(r_{\rm m})}\mathcal{C}(r) L^2(r) \mathrm{d} L ~,
\end{align}
where $L(r,t) \equiv a(t) r e^{\mathcal{R}(r,t)}$ is the so-called areal radius, and $a(t)$ is the scale factor of the background. In the literature, $r_{\rm m}$ is chosen as the radius corresponding to the first local maximum of $\mathcal{C}(r_{\rm m})$.

When $\bar{\mathcal{C}}$ exceeds a universal threshold, often set as $\bar{\mathcal{C}}_{\rm th} = 2/5$, corresponding to $\mathcal{C}_{\rm th} = \mathcal{C}(r_{\rm m})$, this region is believed to collapse into a PBH \cite{Escriva:2019phb}. Using \eqref{eq:compaction function linear}, we can convert $\mathcal{C}_{\rm th}$ to the linear compaction function $\mathcal{C}_{\ell,\rm th}$.
Here, we consider type I perturbations, where $\mathcal{C}_\ell$ takes values in the range $\mathcal{C}_{\ell,\rm th}<\mathcal{C}_\ell<4/3$, corresponding to a monotonically increasing areal radius $L(r)$.

Given a profile of $\mathcal{R}(r)$, using the typical profile from \textit{peak theory} \cite{Bardeen:1985tr,Yoo:2018kvb,Atal:2019cdz,Yoo:2019pma,Atal:2019erb,Yoo:2020dkz,Kitajima:2021fpq}, the type I local maximum $ \mathcal{C}(r_{\rm m})$ can be determined by the condition
\begin{equation}\label{eq:type1 R}
	\mathcal{R}^{\prime} + r\mathcal{R}^{\prime\prime} = 0 ~.
\end{equation}
Here, following the standard approach in the literature \cite{Bardeen:1985tr}, 
and in the limit of high peaks, $\nu = \mu / \sigma_0 \gg 1$ \cite{Atal:2019erb}, we parametrize the profile of the Gaussian curvature perturbation in real space as
\begin{equation}\label{eq:Gaussian typical profile}
	\mathcal{R}_{\mathrm{G}}(r) = \mu \psi_0(r)~,
\end{equation}
where
\begin{equation}\label{eq:mean profile variance}
	\psi_0(r) = \frac{1}{\sigma_0^2} \int\frac{\mathrm{d}k}{k} \frac{\sin(kr)}{kr}\mathcal{P}_{\mathcal{R}_{\mathrm{G}}}(k)~,
	 \qquad 
	 \sigma_0^2 = \int\frac{\mathrm{d}k}{k}\mathcal{P}_{\mathcal{R}_{\mathrm{G}}}(k)~.
\end{equation}
The profile of the actual non-Gaussian curvature perturbation should also relate to the two-point function $\mathcal{R}(r) \propto \expval{\mathcal{R}(r_{0} = 0)\mathcal{R}(r)}$, where $\mathcal{R}(r_{0})$ is the peak of the actual curvature perturbation and we take $r_0 = 0$ as the center of the peak. 

However, the nonlinear relation $\mathcal{R} = \mathcal{F}\left[ \mathcal{R}_{\mathrm{G}}\right]$ implies that $\expval{\mathcal{R}(0)\mathcal{R}(r)} \neq \mathcal{F}\left[\expval{\mathcal{R}_{\mathrm{G}}(0)\mathcal{R}_{\mathrm{G}}(r)}\right]$. 
In this work, we do not adopt the traditional $\mathcal{F}\left[\expval{\mathcal{R}_{\mathrm{G}}(0)\mathcal{R}_{\mathrm{G}}(r)}\right]$ approach to describe the profile of the non-Gaussian random field $\mathcal{R}(\vb*{x})$ in real space. Instead, we introduce a corrected two-point function $\mathcal{P}_{\mathcal{R}}$ for this purpose.  It should be emphasized that the main conclusions of our work are not affected by the specific parametrization of the profile.  Generally, for a highly nonlinear $\mathcal{F}$, it is difficult to express $\mathcal{P}_{\mathcal{R}}$ in terms of $\mathcal{P}_{\mathcal{R}_{\mathrm{G}}}$ analytically. However, since $\mathcal{P}_{\mathcal{R}_{\mathrm{G}}} \ll \order{1}$, we expect that corrections from higher-order non-Gaussianity parameters will be suppressed by $\mathcal{P}_{\mathcal{R}_{\mathrm{G}}}^n$. Thus the correction to the power spectrum of the non-Gaussian field can be calculated perturbatively, following the perturbative local non-Gaussian relation in \eqref{eq:difinition of local NG},

\begin{equation}\label{eq: P_R with correction from NG}
    \begin{aligned}
 \mathcal{P}_{\mathcal{R}}( \vb*{k})&=\mathcal{P}_{\mathcal{R}_{\rm G}}( \vb*{k}) \left( 1 + 3g_{\rm NL}\expval{\mathcal{R}^2_{\rm G}(\vb*{x})} + 6g^2_{\rm NL}\expval{\mathcal{R}^2_{\rm G}(\vb*{x})}^2  + \cdots \right)\\
     &\quad +  \frac{9k^3}{25\pi^2}f^2_{\rm NL}\int\frac{\dd[3]{\vb*{p}}}{(2\pi)^3}P_{\mathcal{R}_{\rm G}}( \vb*{p})P_{\mathcal{R}_{\rm G}}( \vb*{k} - \vb*{p}) \\
       &\quad +  \frac{3k^3}{\pi^2}g^2_{\rm NL}\int\int  \frac{\dd^3 \vb*{l}\dd^3 \vb*{p}}{(2\pi)^6}P_{\mathcal{R}_{\rm G}}( \vb*{p})P_{\mathcal{R}_{\rm G}}( \vb*{l})P_{\mathcal{R}_{\rm G}}(  \vb*{k}- \vb*{l}-\vb*{p}) \\
    &\quad + \cdots ~,
    \end{aligned}
\end{equation}
where the $ \int\frac{\dd^3 \vb*{k}}{(2\pi)^3} P_{\mathcal{R}_{\rm G}}( \vb*{k})e^{i\vb*{k}\cdot\vb*{r}}\equiv \expval{\mathcal{R}_{\mathrm{G}}(0)\mathcal{R}_{\mathrm{G}}(r)}$ defines the dimensionful power spectrum.

Substituting $\mathcal{P}_{\mathcal{R}_{\mathrm{G}}}$ in \eqref{eq:Gaussian typical profile} and \eqref{eq:mean profile variance} with $\mathcal{P}_{\mathcal{R}}$, accounting for the correction from non-Gaussianity, allows us to parameterise the profile of the actual non-Gaussian curvature perturbation. Based on the standard process outlined above, we can determine the threshold $\mathcal{C}_{\rm th}$ to assess whether an overdense region satisfies the conditions for type-I PBH formation. The analytical expression for the threshold given in \cite{Escriva:2020tak} provides a convenient way to determine the threshold $\mathcal{C}_{\rm th}$ in the model revisited in Section \ref{sec:revisiting_up_step}.

The core of the extended Press-Schechter formalism is calculating the probability distribution $\mathbb{P}(\mathcal{C}_\ell)$ for the linear compaction function where $\mathcal{C}_{\ell,\rm th}<\mathcal{C}_\ell<4/3$. 
However, the $\delta N$ formalism in Section \ref{sec:revisiting_up_step} only tells the distribution for $\mathcal{R}$. Let $ X = r\mathcal{R}_{\rm{G}}^{\prime} $, $ Y = \mathcal{R}_{\rm{G}} $, so that $\mathcal{C}_\ell$ can be written as
\begin{equation}\label{eq:non_Gaussianity_path}
	\mathcal{C}_\ell = - \frac43\mathcal{J}(Y)X 
    ~,
\end{equation}
where the Jacobian $ \mathcal{J}(Y)= \dd {\mathcal{R}}/\dd {\mathcal{R}_{\rm{G}}}=\mathcal{F}^\prime(\mathcal{R}_{\rm{G}})$. This relation allows us to determine the probability of $\mathcal{C}_\ell$ taking a certain value by integrating the two-dimensional joint probability distribution $\mathbb{P}(X,Y)$ over $X$ and $Y$,

\begin{equation}\label{eq: PDF of Cl}
    \mathbb{P}(\mathcal{C}_\ell) = \int \dd X \int \dd Y \ \mathbb{P}(X,Y) \delta_\mathrm{D}\left[\mathcal{C}_\ell - \mathcal{C}_\ell(X,Y)\right] ~.
\end{equation}

The two-dimensional joint probability distribution $\mathbb{P}(X,Y)$ of $X$ and $Y$ follows a 2D Gaussian distribution and is given by
\begin{equation}\label{eq:2D Gaussian distribution}
	\mathbb{P}(X,Y) = \frac{1}{2\pi\sqrt{\det(\boldsymbol{\Sigma})}}
 	\exp\left( - \frac{\vb{V}^T\boldsymbol{\Sigma}^{ - 1}\vb{V}}{2} \right) ~,
\end{equation}
where $ \vb{V}^T = (X,Y) $, and the covariance matrix $\boldsymbol{\Sigma}$ is
\begin{equation}
    \boldsymbol{\Sigma} = \begin{pmatrix}\Sigma_{XX}&\Sigma_{XY} \\ \\ \Sigma_{YX}&\Sigma_{YY}\end{pmatrix} ~,
\end{equation}
where each matrix element can be calculated from the power spectrum of curvature perturbations, $\mathcal{P}_{\mathcal{R}_G}(k)$, as given in \cite{Young:2022phe,Gow:2022jfb}:

\begin{equation}\label{eq:covariance elements}
	\begin{aligned}
		\Sigma_{XX}& = \int\dd(\ln k) (kr)^2\left[\frac{\mathrm{d}j_0}{\mathrm{d}z}(kr)\right]^2\mathcal{P}_{\mathcal{R}_{\rm{G}}}(k) W_{\rm TH}^2(k,r)~, \\
		\Sigma_{XY}=\Sigma_{YX}&= \int\mathrm{d}(\ln k) (kr) j_0(kr)\frac{\mathrm{d}j_0}{\mathrm{d}z}(kr) \mathcal{P}_{\mathcal{R}_{\rm{G}}}(k) W_{\rm TH}^2(k,r)~, \\
		\Sigma_{YY}& =  \int\mathrm{d}(\ln k)  j_0^2(kr) \mathcal{P}_{\mathcal{R}_{\rm{G}}}(k) W_{\rm TH}^2(k,r)~,
	\end{aligned}
\end{equation}
where $j_0(z) = \sin(z)/z$, and $W_{\rm TH}(k,r)$ is the top-hat window function in Fourier space. The functional form of $W_{\rm TH}(k,r)$ matches that of the transfer function used in the literature \citep{Choudhury:2023fwk,Young:2024jsu,Iovino:2024tyg,Choudhury:2024aji,Kohri:2024qpd}, and it suppresses large modes ($k \gg 1/r$) when calculating the variance of random fields at a typical scale $r$, 
\begin{equation}
	W_{\rm TH}(k,r) = 3 \frac{\sin\left(kr\right) - (kr)\cos\left(kr\right)}{(kr)^3} ~.
\end{equation}
Note that in the computation of $\mathbb{P}(X,Y)$, we set $r = r_m$.
By substituting \eqref{eq:2D Gaussian distribution} into \eqref{eq: PDF of Cl} and integrating out $X$, $\mathbb{P}(\mathcal{C}_\ell)$ can be rewritten as a single-variable integral over $Y$ \citep{Gow:2022jfb},
\begin{equation}\label{eq:probability_C_l}
	\mathbb{P}(\mathcal{C}_\ell)=\int_{-\infty}^{+\infty}  \frac{3}{4|\mathcal{J}(Y)|} \mathbb{P}\left(-\frac{3}{4}\frac{\mathcal{C}_\ell}{\mathcal{J}(Y)}, Y\right) \dd{Y} ~.
\end{equation}
Using $\mathcal{C}_\ell$, we can express the PBH mass spectrum as follows \cite{Kitajima:2021fpq}:
\begin{equation}
	\beta(M)\mathrm{d}\ln M = \mathcal{K}\frac{\left[\left(\mathcal{C}_\ell - \frac{3}{8}\mathcal{C}_\ell^2\right) - \mathcal{C}_{\mathrm{th}}\right]^{\gamma + 1}}{\gamma\left(1 - \frac{3}{4}\mathcal{C}_\ell\right)}\mathbb{P}(\mathcal{C}_\ell)\operatorname{d}\ln M~,
\end{equation}
which accounts for critical collapse effects \cite{Choptuik:1992jv,Evans:1994pj,Koike:1995jm,Niemeyer:1997mt,Green:1999xm,Musco:2008hv,Musco:2012au,Escriva:2019nsa}:
\begin{equation}
	M(\mathcal{C}_\ell)\sim\mathcal{K}\bigg[\left(\mathcal{C}_\ell - \frac{3}{8} \mathcal{C}_\ell^2\right) - \mathcal{C}_{\mathrm{th}}\bigg]^\gamma M_H~.
\end{equation}
The relation between the current PBH abundance $f(M)$ as a fraction of dark matter and their initial abundance $\beta(M)$ is given by \cite{Kitajima:2021fpq}:
\begin{equation}
	f(M) = \frac{\Omega_{\mathrm{m}}h^2}{\Omega_{\mathrm{DM}}h^2}\frac{T_{r_m}}{T_{\mathrm{eq}}}\beta(M) ~.
\end{equation}

We use the relationship between temperature $T$ and wave number $k$ from \cite{Inomata:2018epa}, and the fitted formulae for the number of degrees of freedom at different temperatures in \cite{Saikawa:2018rcs}, to jointly give the conversion factor between $f(M)$ and $\beta(M)$. The current total PBH abundance as a fraction of dark matter is then given by
\begin{equation}
	f_{\mathrm{PBH}} = \int_0^{M_{\text{max}}} f(M)\mathrm{d}\ln M ~.
\end{equation}
\subsection{PBH abundance}

By utilizing the extended Press-Schechter formalism, we find that non-Gaussianity affects the abundance of PBHs primarily in two aspects, both stemming from the nonlinear relation $\mathcal{R} = \mathcal{F}[\mathcal{R}_{\rm G}]$:
\begin{itemize}
    \item The first effect is on the configuration of the actual curvature perturbation in real space, $\mathcal{R}(r)$, which can deviate from the Gaussian case. As shown in \eqref{eq: P_R with correction from NG}, the two-point function, which mainly determines the profile around the peak, is generally not significantly affected. However, due to the nonlinearity in \eqref{eq:non_linear R}, there is a cut-off for $\mathcal{R}_{\rm G}$, which also imposes a cut-off in the compaction function $\mathcal{C}(r)$. This leads to changes in $r_{\rm m}$ and the threshold $\mathcal{C}_{\rm th}$, thereby altering the probability distribution $\mathbb{P}(X,Y)$.
    \item The second effect arises directly from the non-trivial integration path introduced by the nonlinear relation \eqref{eq:non_Gaussianity_path} when calculating the probability $\mathbb{P}(\mathcal{C}_{\ell})$ using \eqref{eq:probability_C_l}. This path influences how non-Gaussianity impacts PBH abundance through the modified probability space.
\end{itemize}

To understand the effects of the non-Gaussian tail arising from \eqref{eq:non_linear R}, we compute the PBH abundance using the extended Press-Schechter formalism in a concrete model shown in Figure \ref{fig:V_phi}.

\begin{figure}[htbp]
    \centering
    \includegraphics[height=5.5cm]{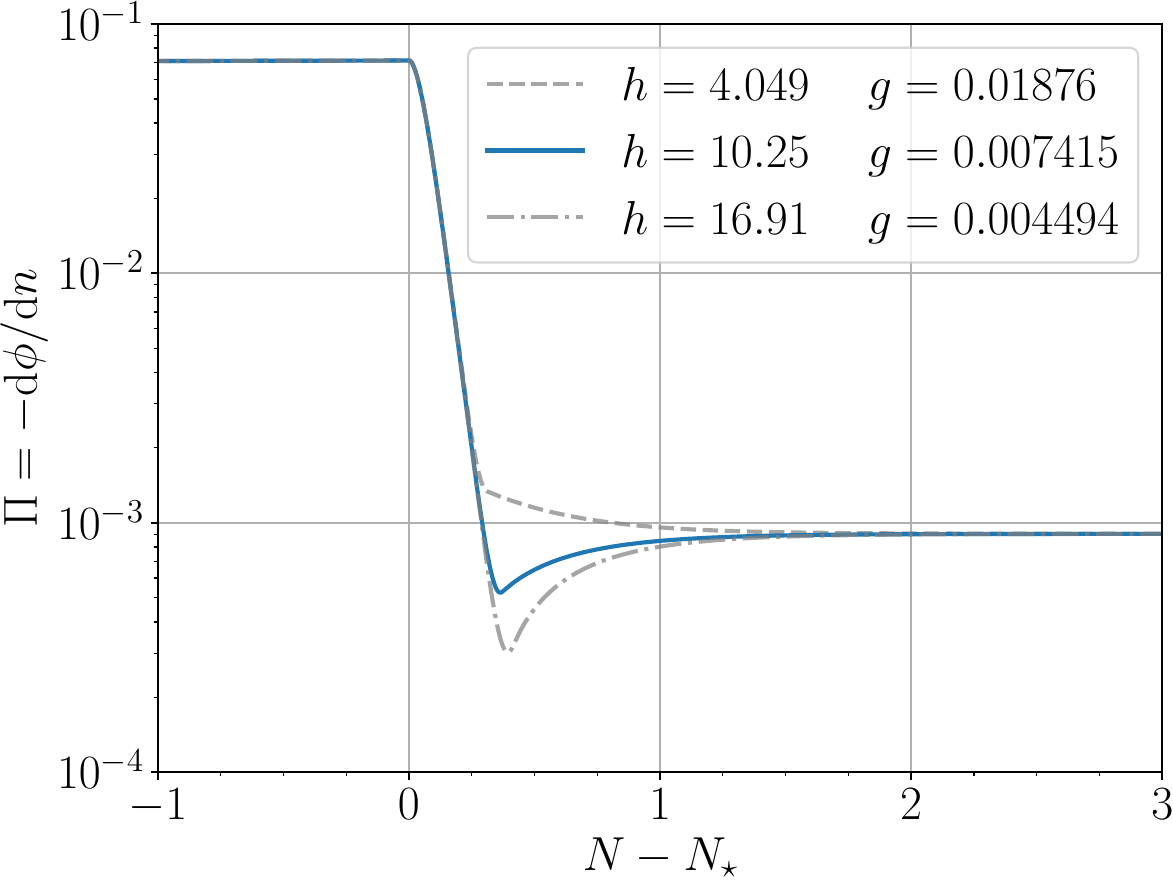}
    \hfill
    \includegraphics[height=5.5cm]{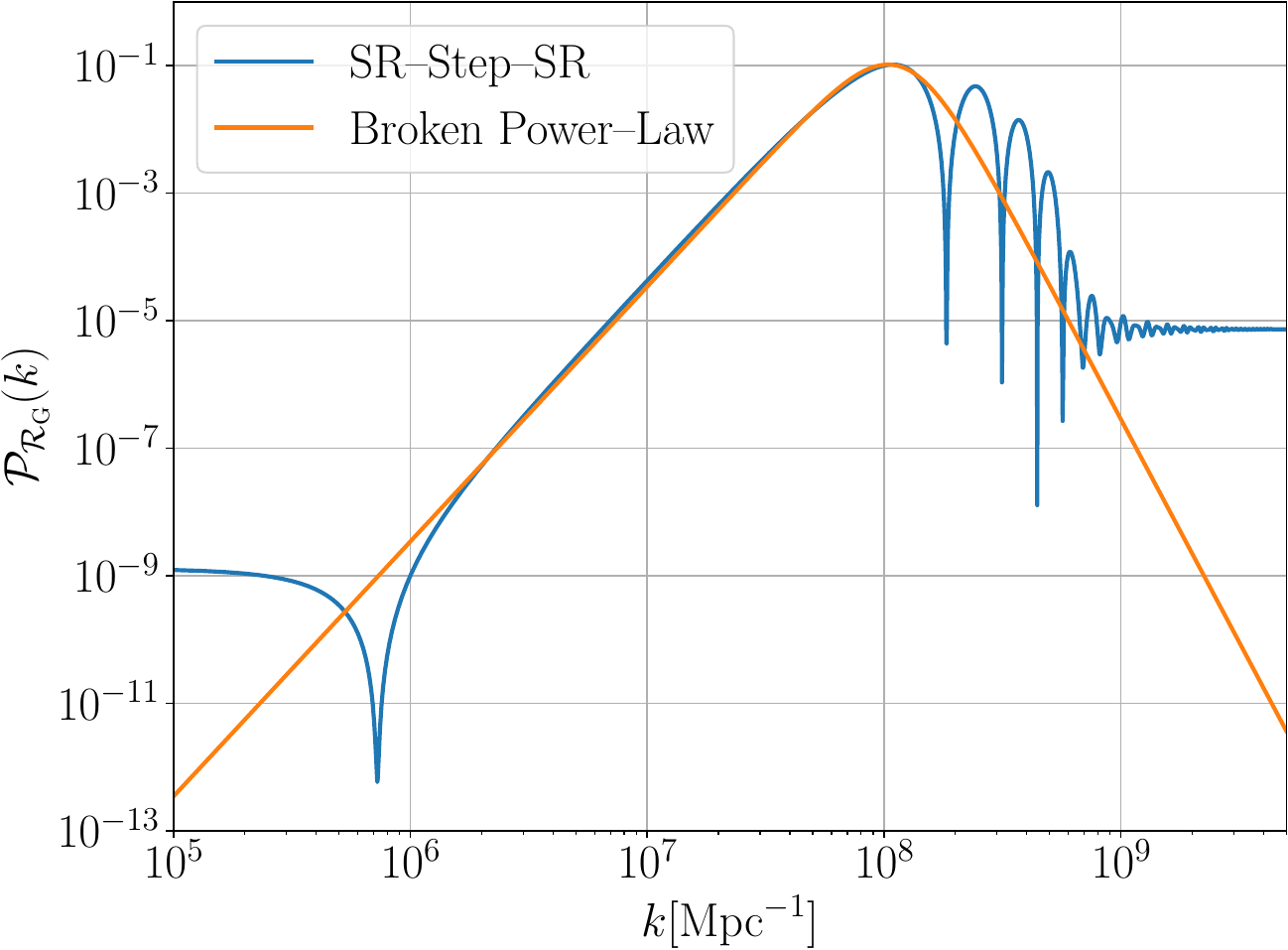}
    \caption{\textbf{Left panel:} The evolution of the inflaton velocity $\Pi(n)$ near the step. The sharp upward step significantly reduces $\Pi(n)$ within $0.5$ e-folds. Changing $\Delta V$ results in different curves, each corresponding to different values of $g$ and $h$. The blue line corresponds to the potential shown in Figure \ref{fig:V_phi}. 
    \textbf{Right panel:} The dimensionless power spectrum $\mathcal{P}_{\mathcal{R}_{\rm{G}}}(k)$ for the model with the potential from Figure \ref{fig:V_phi}. A BPL parametrized by \eqref{eq:ps_bpl} with $A=0.104$, $\alpha=4$, $\gamma=3$, $\beta=7$, and $k_*=1.04\times10^8 \ {\rm Mpc^{-1}}$ fits the result well.}
    \label{fig: pi and power spectrum}
\end{figure}

Using the same parameters as in Figure \ref{fig:V_phi}, one can solve the Sasaki-Mukhanov equation numerically to calculate the power spectrum $\mathcal{P}_{\mathcal{R}_{\rm{G}}}(k)$ at the end of inflation,
\begin{align}    \label{eq:Sasaki-Mukhanov equation}
    \dv[2]{\mathcal{R}_{\rm{G},k}}{\tau}+2\frac{1}{z}\dv{z}{\tau}\dv{\mathcal{R}_{\rm{G},k}}{\tau}+k^2\mathcal{R}_{\rm{G},k}=0,
\end{align}
where $\dd\tau = \dd t/a$ is the conformal time and $z\equiv a \dd\phi/\dd n$. As discussed in Section \ref{sec:revisiting_up_step} and \eqref{eq:power spectrum}, the power spectrum is enhanced due to the rapid decrease in $\Pi$. A broken power-law (BPL) function \eqref{eq:ps_bpl} can effectively describe the peak behavior of $\mathcal{P}_{\mathcal{R}_{\rm{G}}}(k)$, as shown in Figure \ref{fig: pi and power spectrum}. 

Next, we will estimate the PBH abundance in this model based on the BPL parameterized power spectrum,
\begin{equation}\label{eq:ps_bpl}
	\mathcal{P}_{\mathcal{R}_{\rm{G}}}^\mathrm{BPL}(k) = A\frac{(\alpha + \beta)^\gamma}{\left[\beta\left(k/k_*\right)^{ - \alpha/\gamma} + \alpha\left(k/k_*\right)^{\beta/\gamma}\right]^\gamma} ~.
\end{equation}
Since PBH formation is primarily determined by the peak of the power spectrum, this approximation does not affect the results of our analysis much.

By employing the extended Press-Schechter formalism outlined in Section \ref{sec:Basic setup for extended Press-Shecheter formalism}, we find that the abundance of PBHs from the type-I channel is significantly suppressed by the non-Gaussian tail described by \eqref{eq:non_linear R}. To understand how the non-Gaussian tail, controlled by $h$, affects $f_{\rm PBH}$, we compute $f_{\rm PBH}$ for different values of $h$, keeping $\mathcal{P}_{\mathcal{R}_{\rm{G}} }(k)$ fixed to the BPL spectrum shown in Figure \ref{fig: pi and power spectrum}. And we perturbatively estimate the non-Gaussian correction to $\mathcal{P}_{\mathcal{R}}(k)$ by using \eqref{eq: P_R with correction from NG} as shown in Figure \ref{fig:nG PR}:
\begin{figure}
    \centering
    \includegraphics[width=0.7\linewidth]{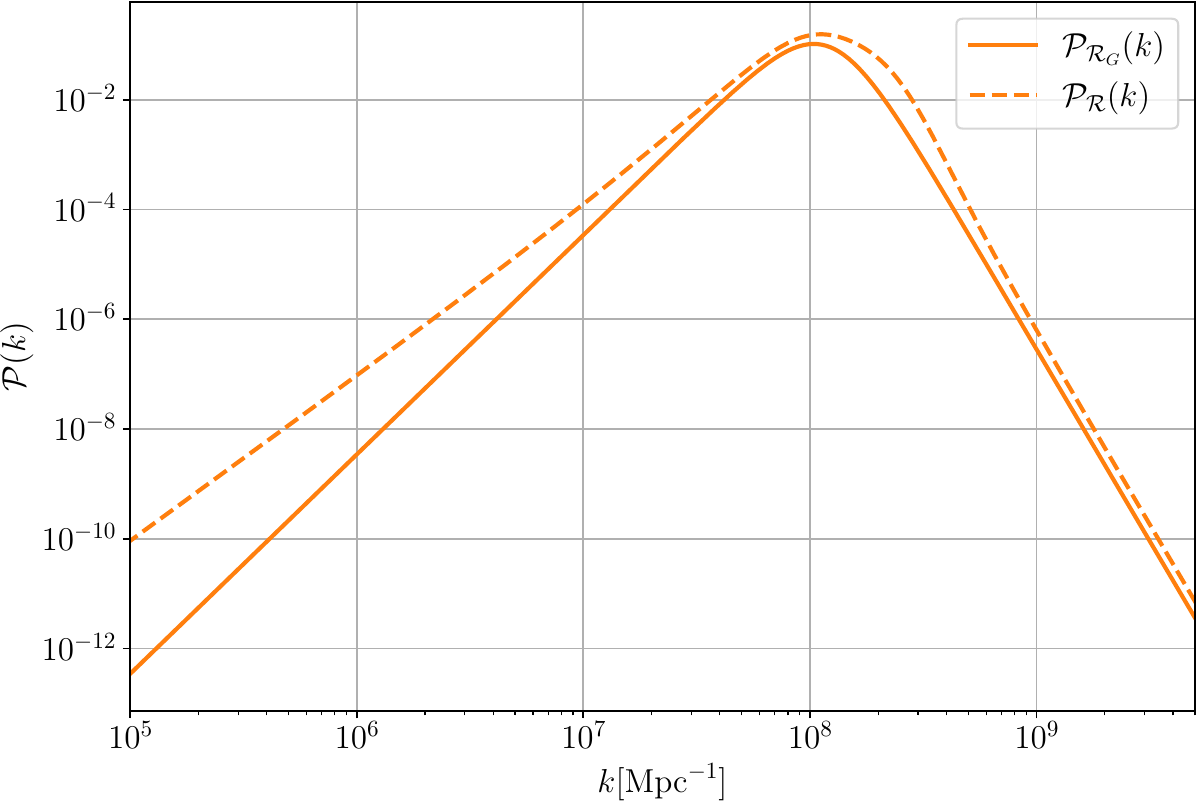}
    \caption{The comparison of $\mathcal{P}_{\mathcal{R}}(k)$ with non-Gaussian correction from   \eqref{eq: P_R with correction from NG} and BPL $\mathcal{P}_{\mathcal{R}_{\rm{G}} }(k)$ from the right panel of Figure \ref{fig: pi and power spectrum}.}
    \label{fig:nG PR}
\end{figure}

\begin{figure}[!htbp]
    \centering
    \includegraphics[width=0.8\linewidth]{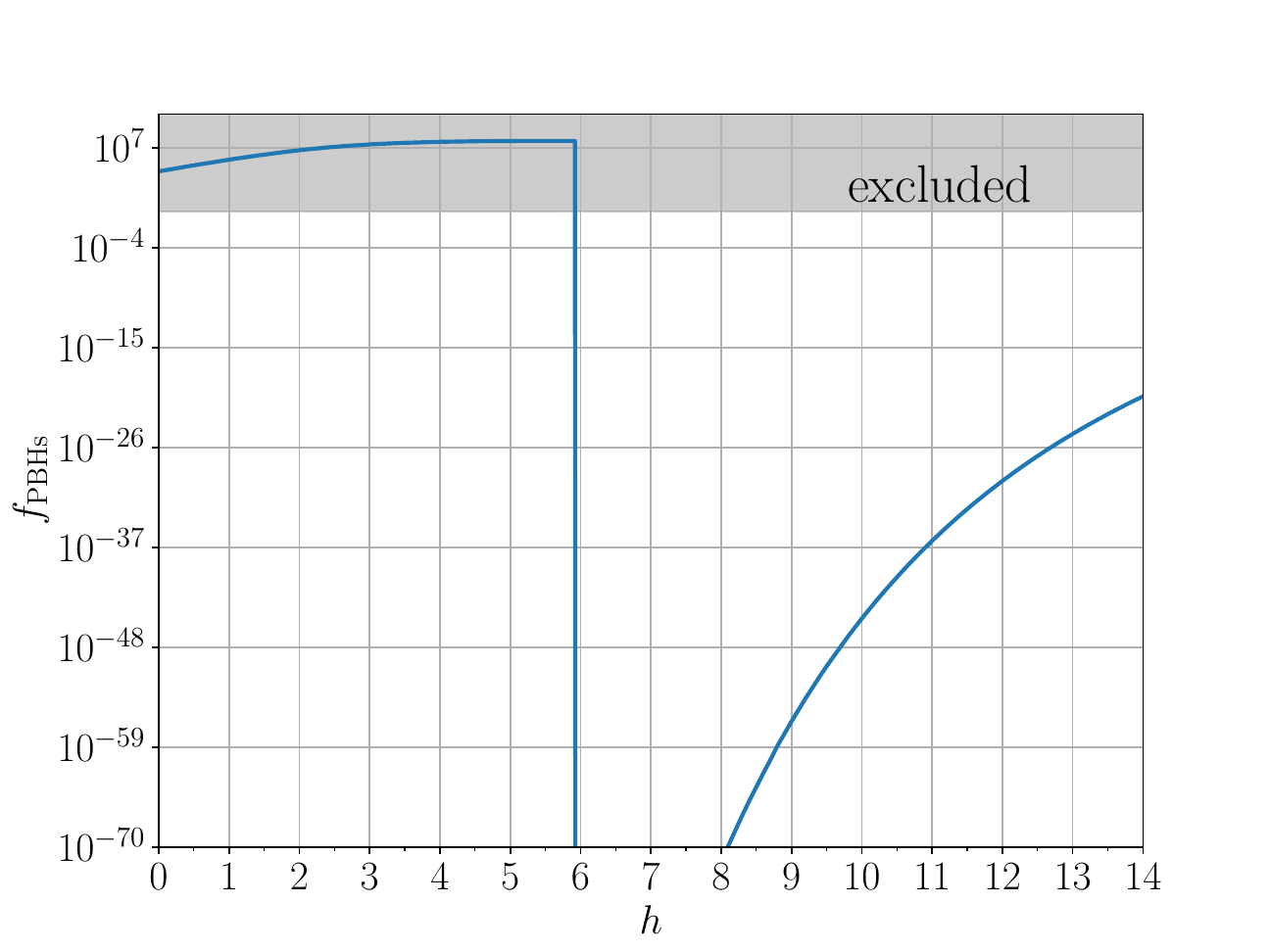}
    \caption{With the given BPL power spectrum, the abundance of PBHs from the type-I channel is evaluated for various $h$ values, considering the impact of the non-Gaussian tail. When $h < 5.9 $, the non-Gaussianity introduced by \eqref{eq:non_linear R} enhances PBH formation. However, when $h > 5.9 $, the cutoff effect of \eqref{eq:non_linear R} leads to a sharp reduction in the PBH abundance from the type-I channel. As $h$ increases further, $f_{\rm PBH}$ begins to rise again, but at this point, a more precise calculation for the type-II channel is likely necessary.}
    \label{fig:fPBH_vs_h}
\end{figure}

As shown in Figure \ref{fig:fPBH_vs_h}, as $h$ increases from $h = 0$, $f_{\rm PBH}$ is initially enhanced by the non-Gaussian tail compared to the Gaussian case ($h = 0$). For the $\mathcal{P}_{\mathcal{R}_{\rm{G}}}(k)$ in Figure \ref{fig: pi and power spectrum}, PBHs are overproduced in the Gaussian scenario. However, as $h$ increases beyond approximately $h \simeq 5.9$, $f_{\rm PBH}$ drops by a factor of $10^{133}$ and then begins to rise again. 

The sharp drop in $f_{\rm PBH}$ around $h \simeq 5.9 $ is not unexpected. We can explain this, even with a rough estimate, as anticipated in previous studies \cite{Cai:2022erk}. Since the averaged density contrast is proportional to the curvature perturbation at horizon crossing, $\mathcal{R}_{\rm th} \sim \mathcal{O}(1)$ serves as a rough criterion for PBH formation. However, the non-linear relation \eqref{eq:non_linear R} implies a cut-off at $2/h$ in the PDF of $\mathcal{R}$. Using the extended Press-Schechter formalism, we confirm this naive argument.

From the perspective of extended Press-Schechter formalism, the sudden drop of $f_{\rm PBH}$ can be seen as the evidence for type-II fluctuation of PBHs. We show the profile of compaction function $\mathcal{C}(r)$ of non-Gaussian curvature perturbation $\mathcal{R}(r)$ utilized by different $\mu$ in Section \ref{sec:Basic setup for extended Press-Shecheter formalism} with different $h$ to understand it clearly.
\begin{figure}[htbp]
	\centering
	\begin{minipage}{0.485\linewidth}
		\centering
		\includegraphics[width=1\linewidth]{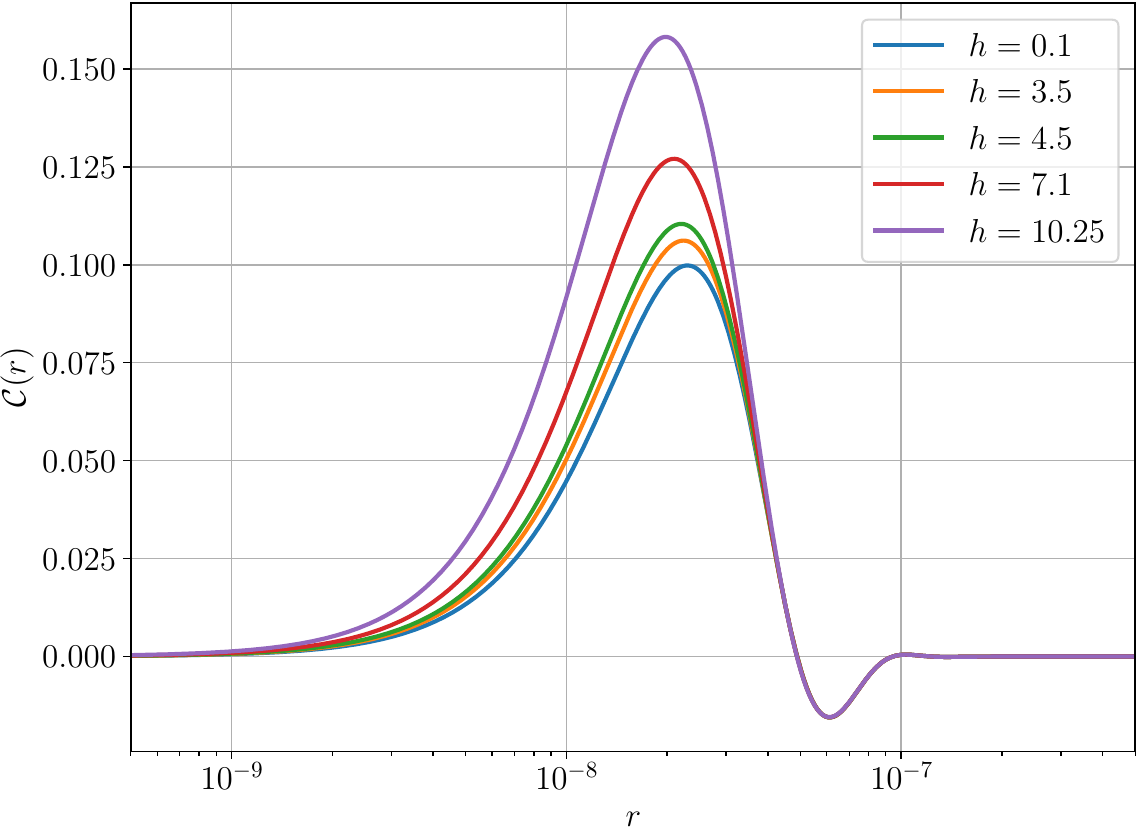}\\
		\text{$\mu_2=0.10$}
	\end{minipage}
 \hfill
	\begin{minipage}{0.485\linewidth}
		\centering
		\includegraphics[width=1\linewidth]{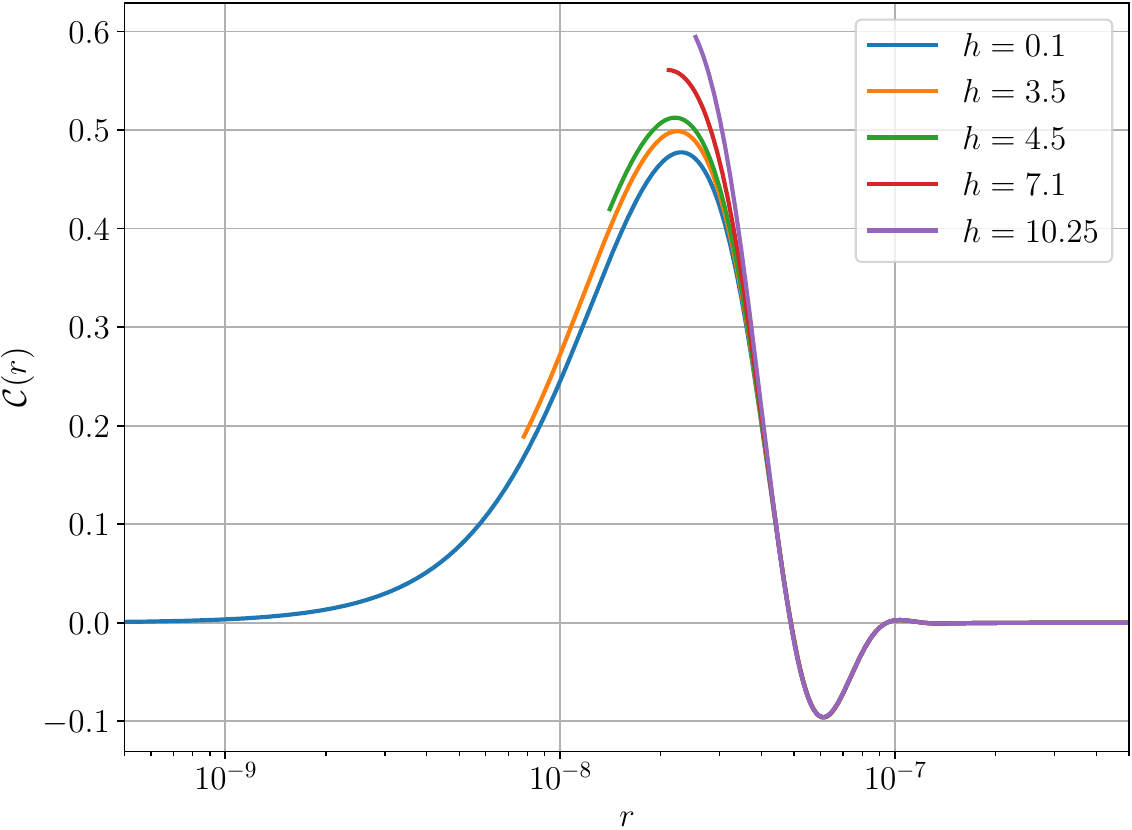}
		\text{$\mu_2=0.60$}
	\end{minipage}
	\caption{
The $\mathcal{C}(r)$ profiles for different $h$ values with two given $\mu$ values. We increase $\mu$ to determine the threshold $\mathcal{C}_{\rm th}$ and $r_{\rm m}$.}\label{fig:diff_mu_diff_h_C_r}
\end{figure}
As discussed in Section \ref{sec:Basic setup for extended Press-Shecheter formalism}, we increase $\mu$ from a small initial value to find $\mathcal{C}_{\rm th} = \mathcal{C}(r_{\rm m})$, where $\mathcal{C}$ reaches a local maximum at $r_{\rm m}$. This $r_{\rm m}$ also determines the joint PDF $\mathbb{P}(X,Y)$ via \eqref{eq:covariance elements}. The nonlinear relation \eqref{eq:non_linear R} introduces a cut-off for $\mathcal{R}_{\rm G}$, leading to a cut-off and a corresponding type-II peak in $\mathcal{C}(r)$ when $\mu$ is large. 

As shown in Figure \ref{fig:diff_mu_diff_h_C_r}, when $\mu$ is small, the first peak in $\mathcal{C}(r)$ corresponds to a standard type-I peak, consistent with \eqref{eq:type1 R}. However, for larger values of $\mu$, $\mathcal{C}(r)$ with larger $h$ results in a cut-off, as seen in the right panel of Figure \ref{fig:diff_mu_diff_h_C_r}. When this cut-off eliminates the first type-I peak, the next type-I peak can be found at larger $r$. 

\begin{figure}
    \centering
    \includegraphics[width=0.7\linewidth]{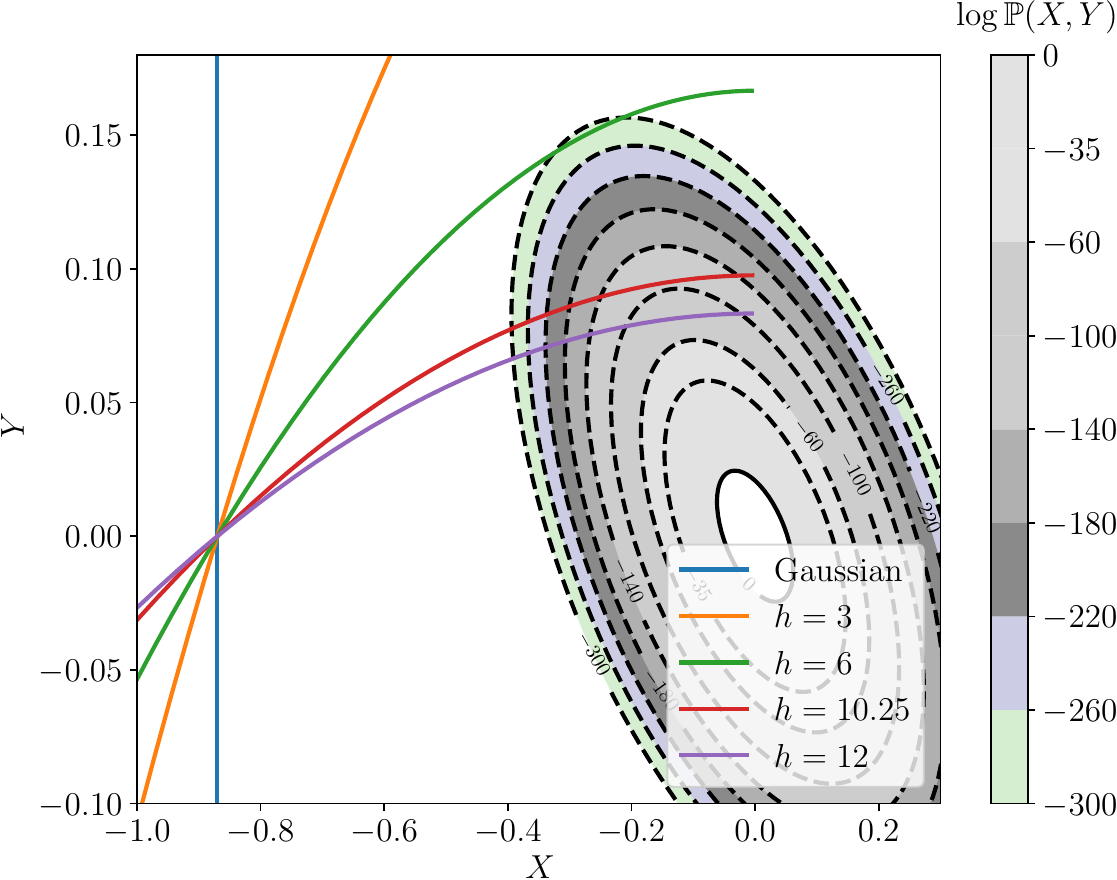}
    \caption{The integration paths for calculating $\mathbb{P}(\mathcal{C}_{\ell})$ corresponding to different $h$ values, with $\mathcal{C}_\ell = 1.16$. When the impact of non-Gaussianity on $\mathbb{P}(X,Y)$ is small, we can roughly estimate the PBH production by considering different integration paths in \eqref{eq:non_Gaussianity_path}. The PDF $\mathbb{P}(X,Y)$ shown in the figure corresponds to $h = 10.25 $.}
    \label{fig:upwrad_step_h_P_2D}
\end{figure}

In practical calculations using the extended Press-Schechter formalism, we find next type-I local maximum, causing $r_{\rm m}$ to increase discontinuously. In this case, PBH formation via the type-I channel is significantly suppressed. For larger $h$, which corresponds to a smaller cut-off in $\mathbb{P}[\mathcal{R}]$, the sudden change in $r_{\rm m}$ occurs with a smaller $\mu$. After this sudden shift, the $r_{\rm{m}}$ corresponding to the type-I peak becomes almost insensitive to changes in $h$. Therefore, we can approximately assume that after the abrupt change in $r_{\rm m}$, further increases in $h$ have little impact on $\mathbb{P}(X,Y)$. 
However, different $h$ values also affect the integral path in \eqref{eq:probability_C_l}. For a given PDF $\mathbb{P}(X,Y)$, larger $h$ shifts the integral path through regions of higher probability, as shown in Figure \ref{fig:upwrad_step_h_P_2D}. This explains the increasing $f_{\rm PBH}$ in Figure \ref{fig:fPBH_vs_h} when $h > 5.9 $. However, it is important to emphasize that when $h > 5.9 $, the sudden drop in $f_{\rm PBH}$ indicates that PBHs formed via type-I fluctuations are suppressed, while the type-II channel may become more significant in this regime \cite{Escriva:2023uko,Shimada:2024eec}.

Estimating the formation of PBHs from Type-II fluctuations is challenging not only due to the complexity arising from the non-monotonic behavior of the areal radius $L(r)$, but also because the mass associated with these fluctuations can be significantly larger than that of the Hubble patch. For modes exiting the horizon just before the step, there is a possibility that the inflaton becomes trapped at the bottom of the step due to large quantum fluctuations. This phenomenon directly leads to the cut-off for $\mathcal{R}_{\rm G}$ in \eqref{eq:non_linear R}. 

Nevertheless, we can approximately estimate the probability of the inflaton being trapped as
\begin{align}
    \mathbb{P}\left(\mathcal{R}_{\rm G} > \frac{1}{h}\right) = \int_{1/h}^{+\infty} \frac{1}{\sqrt{2\pi \sigma_{\mathcal{R}_{\rm G}}^2}}\exp\left(-\frac{\mathcal{R}_{\rm G}^2}{2\sigma_{\mathcal{R}_{\rm G}}^2}\right) \mathrm{d} \mathcal{R}_{\rm G} ~,
\end{align}
where $\sigma_{\mathcal{R}_{\rm G}}^2 = \Sigma_{YY}$, as derived in Section \ref{sec:Basic setup for extended Press-Shecheter formalism}. For the parameters used in the right panel of Figure \ref{fig: pi and power spectrum}, this probability is extremely low, of the order $\mathcal{O}(10^{-54})$.

The model used in Figure \ref{fig:V_phi}, which generates the power spectrum shown in the right panel of Figure \ref{fig: pi and power spectrum}, does not feature an infinitely sharp upward step. Only in the case of an extremely sharp step 
 \eqref{eq:pid} can hold, but in such cases, the power spectrum $\mathcal{P}_{\mathcal{R}_{\rm{G}}}(k)$ would oscillate and fail to decay to the second slow-roll solution. For a step with finite width, \eqref{eq:pid} needs to be modified to account for the Hubble friction as the inflaton crosses the step \cite{Kawaguchi:2023mgk}, although the nonlinear relationship between $\Pi_{\rm c}$ and $\Pi_{\rm d}$ remains. 

\begin{figure}
    \centering
    \includegraphics[width=0.63\linewidth]{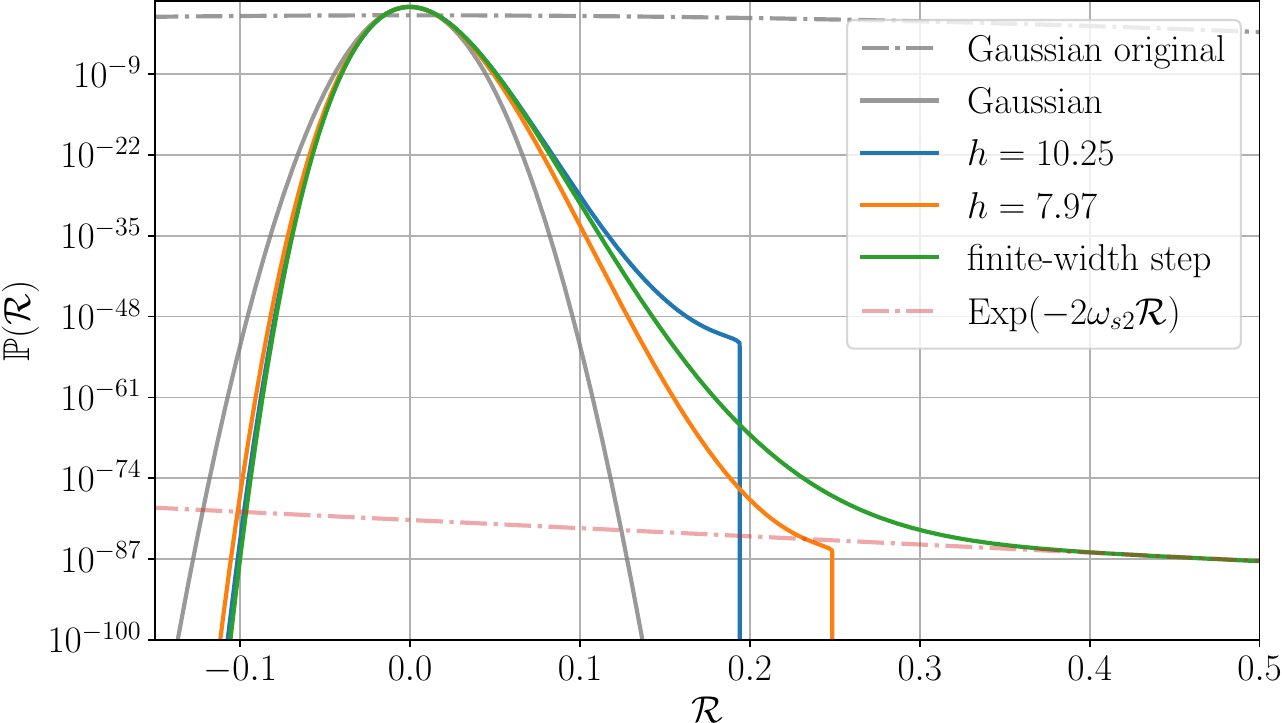}
    \caption{The PDF $\mathbb{P}[\mathcal{R}]$ for modes exiting the horizon just before the upward step is shown. The green line exhibits an exponential tail with $\omega_{s2} \simeq 15.13$, accounting for the finite size effects of the upward step \cite{Kawaguchi:2023mgk}. Two Gaussian distributions are present in this figure: the variance of the broader  one corresponds precisely to the variance of $\mathcal{R}_{\rm G}$ as described in \eqref{eq:non_linear R}. However, as discussed earlier, when  cut-off eliminates the first type-I peak, a sudden increase in $r_{\rm m}$ modifies the variance $\sigma_{\mathcal{R}_{\rm G}}^2$. The variance of the narrower Gaussian is determined by this new $r_{\rm m}$. }
    \label{fig:P_R_step_all}
\end{figure}

As shown in Figure \ref{fig:P_R_step_all}, the finite width of the step removes the hard cut-off in the PDF of $\mathcal{R}$, instead producing an exponentially decaying tail $\mathbb{P}[\mathcal{R}] \propto \exp(-2\omega_{s2}\mathcal{R})$, where the index $\omega_{s2} \simeq \sqrt{2} \; \Pi_{c} /\Delta \phi$ is determined by the step width. In our case, we find $\omega_{s2} \simeq 15.13$, which does not alter the key conclusions. This exponential tail results in an ``equivalent'' $10.25 > h > 7.97$, meaning that the abundance of type-I PBHs from the realistic model without a cut-off is the same as with a modified $10.25 > h > 7.97$.

Additionally, we find that the abundance of PBHs in models with an upward step is highly sensitive to the parameters. A wider step or even a change in the shape of the step can significantly affect the results \cite{Animali:2022otk,Escriva:2023uko}.

\section{Implications for indirect observations of PBHs}\label{sec_PTA_PBHs_overp}

Observing the primordial power spectrum through gravitational waves and CMB $\mu$-distortion to constrain PBH abundance is considered an indirect method. Since the abundance of PBHs in the presence of non-Gaussianity cannot be fully determined by the power spectrum, it introduces significant uncertainty to such indirect observations.

\subsection{Energy density spectrum of SIGWs}\label{subsec_GWs_cal}

For instance, based on PTAs observations of the stochastic gravitational wave background \cite{NANOGrav:2023gor,NANOGrav:2023hde,EPTA:2023fyk,EPTA:2023sfo,EPTA:2023xxk,Reardon:2023gzh,Zic:2023gta,Reardon:2023zen,Xu:2023wog}, if we assume that the entire stochastic background is generated by primordial scalar perturbations, as shown in the right panel of Figure \ref{fig: pi and power spectrum}, the observations imply that $\mathcal{P}_{\mathcal{R}_{\rm G}}(k)\sim \order{10^{-1}}$. At the same time, Figure \ref{fig:fPBH_vs_h} shows that PBHs are overproduced in the Gaussian case ($h = 0$). 

When examining the impact of non-Gaussianity on PBH abundance, a polynomial series expansion \eqref{eq:difinition of local NG} usually indicates that a positive $f_{\rm NL}$ always increases the abundance of PBHs. This reasoning suggests that models with an upward step are disfavored by PTAs observations since they exacerbate the overproduction of PBHs \cite{Firouzjahi:2023xke}. 

\begin{figure}[htbp]
    \centering
    \includegraphics[width=0.76\linewidth]{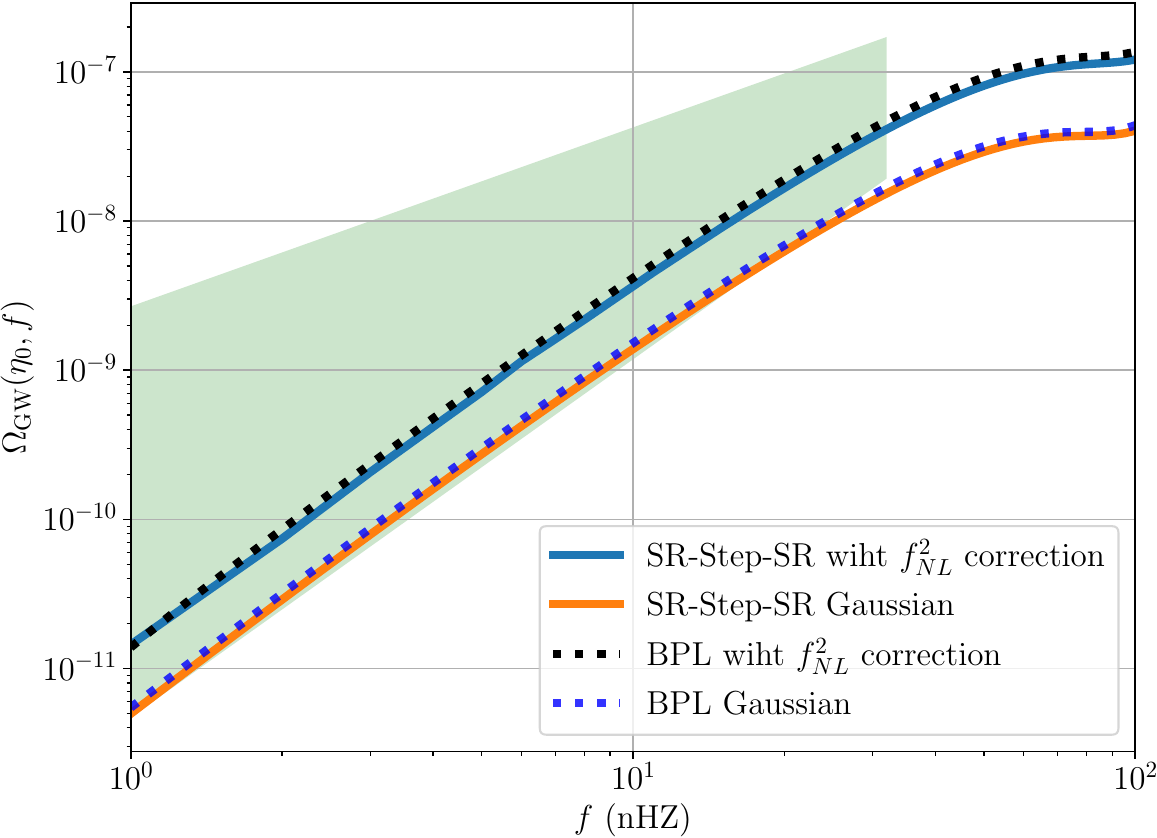}
    \caption{Based on our model, we compute the energy spectrum $\Omega_{\mathrm{GW}}(\eta_0,k)$ using the power spectrum shown in the right panel of Figure \ref{fig: pi and power spectrum}. Additionally, we account for the correction from $f^2_{\rm NL}$ to $\Omega_{\mathrm{GW}}(\eta_0,k)$. The green shaded regions represent the $2\sigma$ confidence interval as indicated by the NANOGrav 15-year data.}
    \label{fig:SIGWs_upstep_and_BPL}
\end{figure}

However, when we consider the effect of non-perturbative non-Gaussianity \eqref{eq:non_linear R}, we find that the issue of overproduction can be resolved, although fine-tuning is required.
Using the power spectrum from the right panel of Figure \ref{fig: pi and power spectrum}, we apply the standard approach to compute the present-day energy spectrum of SIGWs $\Omega_{\mathrm{GW}}(\eta_0,k)$ \cite{Baumann:2007zm,Espinosa:2018eve,Kohri:2018awv,Domenech:2021ztg}, and include corrections up to $f^2_{\rm NL}$ order due to non-Gaussianity \cite{Cai:2018dig,Atal:2021jyo,Adshead:2021hnm,Abe:2022xur,Li:2023qua}. As shown in Figure \ref{fig:SIGWs_upstep_and_BPL}, the SIGWs produced in the upward step model align well with the NANOGrav 15-year data without leading to an overproduction of PBHs.

\section{Conclusion}\label{sec:conclusion}

In this work, we have investigated the formation of PBHs within the framework of an upward step inflationary model. A key finding is that the nonlinear relation between the curvature perturbation and the field fluctuation, $\mathcal{R} (\delta \phi)=\mathcal{F}[\mathcal{R}_{\rm G}]$, introduces a cutoff that deviates significantly from the Gaussian case, necessitating a reevaluation of PBH formation mechanisms. This necessitates a careful reevaluation of PBH formation mechanisms, as the curvature perturbation $\mathcal{R}$ is not the appropriate variable for calculating the abundance of PBHs \cite{Young:2014ana,DeLuca:2022rfz,Raatikainen:2023bzk}.

By employing the extended Press-Schechter formalism, we demonstrate that non-Gaussianity significantly influences PBH formation through the nonlinear relation between the curvature perturbation and its Gaussian counterpart. This nonlinearity introduces a cut-off in the curvature perturbation, which impacts PBH formation in two key ways: 

First, it modifies the compaction function $\mathcal{C}(r)$, leading to changes in the characteristic scale $r_{\rm m}$, the threshold $\mathcal{C}_{\rm th}$, and the probability distribution $\mathbb{P}(X,Y)$. Second, the nonlinear relation affects the integration path used to compute the probability $\mathbb{P}(\mathcal{C}_{\ell})$, altering the estimation of PBH abundance by influencing the regions of integration in probability space. Consequently, the formation of PBHs, particularly through the type-I channel, is highly sensitive to this non-Gaussian correction.

Our results show that as the non-Gaussian parameter $h$ increases, the PBH fraction $f_{\rm PBH}$ initially increases compared to the Gaussian case. However, beyond a critical value of $h \simeq 5.9$, $f_{\rm PBH}$ sharply declines before gradually increasing again. This behavior is due to the cutoff in the probability distribution function of $\mathcal{R}$ induced by the nonlinear relation, effectively suppressing PBH formation via the type-I channel when $h$ exceeds the critical value.

This study has significant implications for indirect observations of PBHs through gravitational waves and CMB $\mu$-distortions. In the presence of non-Gaussianity, the abundance of PBHs cannot be fully determined by the power spectrum alone, introducing significant uncertainty to such indirect observations. For instance, based on PTAs observations, if we assume that the entire stochastic gravitational wave background is generated by primordial scalar perturbations, the observations imply a large amplitude of $\mathcal{P}_{\mathcal{R}_{\rm{G}}}(k)\sim \order{10^{-1}}$. In the Gaussian case, this would lead to an overproduction of PBHs. However, considering the effect of non-perturbative non-Gaussianity, the overproduction issue can be resolved, aligning the model with current observations without conflicting with PBH constraints.

Our findings highlight that strong non-Gaussian features can significantly impact the interpretation of indirect PBH observations. This underscores the necessity of incorporating non-Gaussian effects into theoretical models and observational analyses to accurately predict PBH abundances and interpret gravitational wave data.

In conclusion, this study emphasizes the crucial role of non-Gaussianity in PBH formation and its implications for indirect observational constraints. The upward step model demonstrates how non-Gaussian features can mitigate the overproduction of PBHs without conflicting with current observations. In future work, investigation of type-II fluctuations in such models using numerical simulation methods may provide deeper insights into PBH formation mechanisms and their observational signatures.

\begin{acknowledgments}
We thank Albert Escrivà, Gabriele Franciolini, Xin-Chen He, Kaloian Dimitrov Lozanov, Shi Pi, Misao Sasaki, Xinpeng Wang, Qiyue Xia, Shuichiro Yokoyama, Chul-Moon Yoo, and Dongdong Zhang for valuable discussions.
This work is supported in part by the National Key R\&D Program of China (2021YFC2203100), by NSFC (12261131497), by CAS young interdisciplinary innovation team (JCTD-2022-20), by 111 Project (B23042), by Fundamental Research Funds for Central Universities, by CSC Innovation Talent Funds, by USTC Fellowship for International Cooperation,  by USTC Research Funds of the Double First-Class Initiative, by the National Key Research and 
Development Program of China (2021YFC2203203), and by the Science and Technology Plan Projects of the Tibet Autonomous Region (XZ202102YD0029C). It is also supported in part by the JSPS KAKENHI grants No. 20H05853 and 24K00624.

We would also like to express our sincere gratitude to the Dynamics of Primordial Black Hole Formation II (DPBHF2) workshop held in Nagoya, which provided valuable insights and greatly contributed to the progress of this work.

\end{acknowledgments}

\bibliographystyle{JHEP}
\bibliography{biblio.bib}

\end{document}